\documentclass{vgtc}

\ifpdf                               
  \pdfoutput=1\relax                   
  \usepackage{graphicx}              
  \DeclareGraphicsExtensions{.pdf,.png,.jpg,.jpeg} 
\else%
  \usepackage{graphicx}
  \usepackage{wrapfig}
  \usepackage{picinpar}
  \usepackage{epstopdf}
  \DeclareGraphicsExtensions{.eps}     
\fi

\graphicspath{{figures/}{pictures/}{images/}{./}}

\usepackage{microtype}                 
\PassOptionsToPackage{warn}{textcomp}  
\usepackage{textcomp}                  
\usepackage{mathptmx}                 
\usepackage{times}                    
        
\usepackage{cite}                      
\usepackage{tabu}                      
\usepackage{booktabs}  
\usepackage{enumitem}
\usepackage{graphicx}
\usepackage{caption}
\usepackage{subfigure}
\usepackage{url}
\onlineid{0}
\vgtccategory{Research}
\vgtcinsertpkg

\title{A Study on Activity Visualization for Smart Watches }

\author{Zhouxuan Xia\thanks{e-mail: \parbox{1.8cm}{Zhouxuan.Xia21@student.xjtlu.edu.cn}}
\and Yu Liu\thanks{Corresponding author. e-mail: \parbox{1.8cm}{Yu.Liu02@xjtlu.edu.cn}}
\and Fabiola Polidoro\footnotemark[1] \thanks{e-mail: \parbox{1.8cm}{F.Polidoro18@student.xjtlu.edu.cn}}
}
\affiliation{\scriptsize Xi’an Jiaotong-Liverpool University, Suzhou, China}

\vspace{-2.5cm}

\abstract{This paper investigates the use of visualization to display activity data on smartwatches by surveying the data visual presentations proposed by 80 smartwatch models currently available on the Chinese e-commerce platform JD.com and, later, surveying the preferences of 41 users concerning these visualizations. The results show that despite radial bar charts are the most popular visualization for activity data on smartwatches, the users' preferences might be influenced by their familiarity with these charts. These findings from this survey will be valuable for designers, developers, and researchers who are interested in creating innovative and effective solutions for activity visualization on smartwatches.}

\CCScatlist{
  \CCScatTwelve{Human-centered computing}{Visu\-al\-iza\-tion}{Visualization design and evaluation methods}{}
}

\begin{document}

\maketitle

\section{Introduction}
Nowadays, smartwatches harvested a lot of consumers worldwide -- they are expected to register a Compound Annual Growth Rate (CAGR) of 21.98$\%$ from 2021 to 2026\cite{WebCatation} -- and have become an indispensable part of our daily life by revolutionized the way we interact with technology. With the integration of sensors such as GPS, gyroscopes and accelerometers, smartwatches are capable of tracking various physical activities, e.g. running, walking, and swimming, display the collected data as micro visualizations and even provide the user with recommendations for improving their life style \cite{angelides2018wearable}.

Previous studies have explored various aspects of activity visualization for smartwatches, including the representation of health and fitness data \cite{Schiewe2020ASoR}, methods for filtering and highlighting notifications \cite{Islam2020VIoW}, glanceable feedback for group fitness activities \cite{8443125}, visual representations on  activity tracking data \cite{Schiewe2020ASoR}, comparison on sleep data \cite{Islam2022PaEo} . Some studies have even proposed novel visualizations, such as real-time feedback for runners \cite {Schiewe2020ASoR},  personal wearable data in immersive environments\cite{angelides2018wearable} and visual displays in the sightseeting context \cite{islam2022context}. Another study presented recommendations generated through machine learning techniques on activity data visualization\cite{angelides2018wearable}. These studies provide important insights into the field of visualization for smartwatches.
With this work, we aim at expanding the previous studies on smartwatch visualization and user preferences with a focus on the Chinese market, by presenting the results of 1) a survey of the visualizations employed by 80 best-selling smartwatch models on the Chinese e-commerce platform JD.com and 2) a survey of the visualization preferences and interests of 42 smartwatch users.
\vspace{-1mm}

\section{ACTIVITY DATA ON SMARTWATCHES}
For the first exploratory study we surveyed the popular Chinese e-commerce platform for electronic devices, JD.com, and analyzed the most used visualizations for activity tracking data (namely the time spent standing, moving and exercising) among the best-selling smartwatches. This resulted in 80 smartwatch models, belonging to the following brands: Apple (25), Huawei (17), Mi (2), Oppo (4), Vivo (1), Honor (1), Samsung (6), Dido (7), Garmin (7), Fitbit (5) and Amazfit (5). We classified the visualizations displaying activity data according to the screen shape of the watch, the visualization type, the arrangement of the data, and the number of different visualizations displayed at once on the watch face.

\begin{figure}[h]
    \begin{minipage}[c]{0.4\linewidth}
        \includegraphics[width=\textwidth]{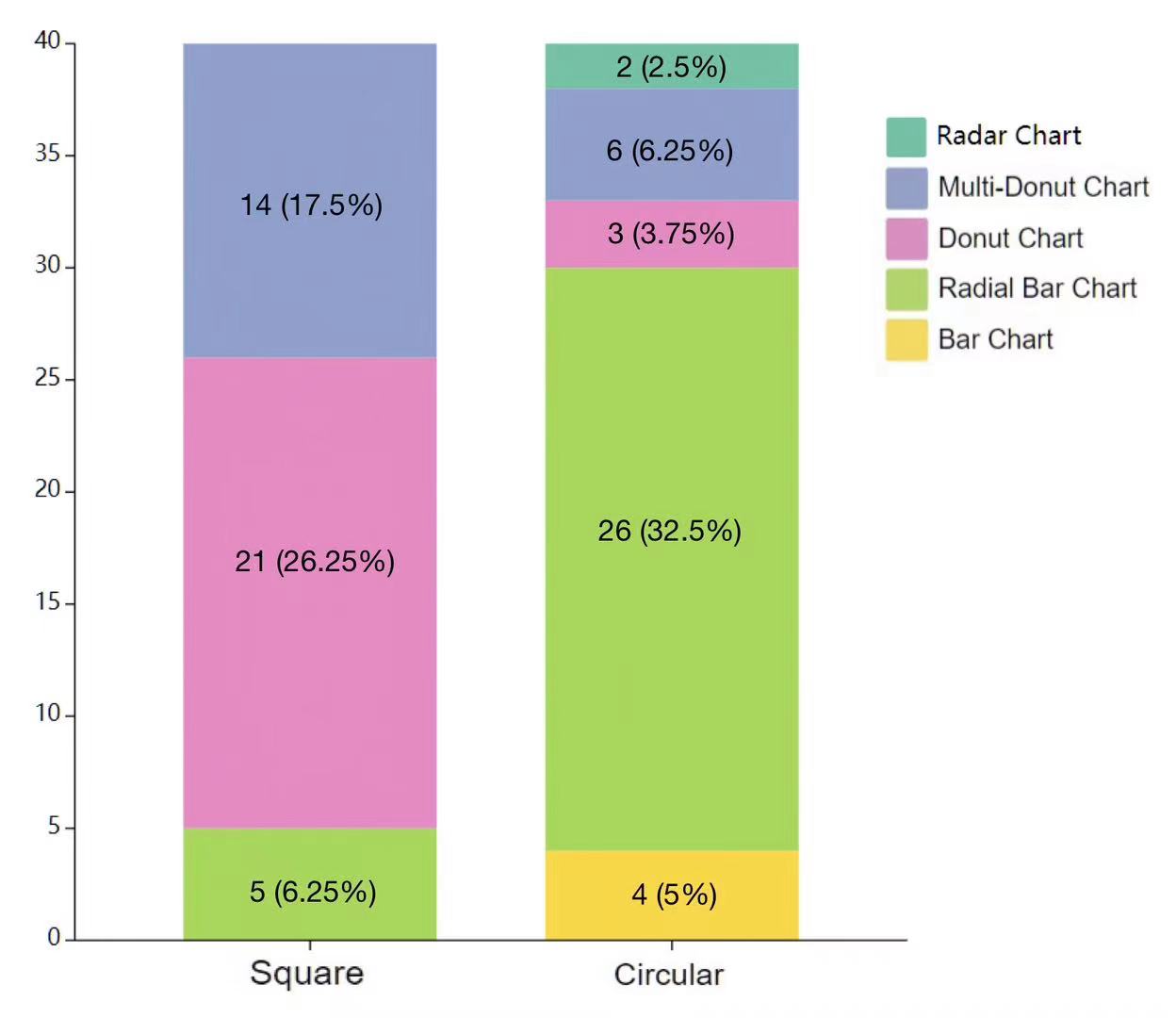}
    \end{minipage}\hfill
    \begin{minipage}[c]{0.50\linewidth}
        \caption{
           \emph{Visualizations on square and circular smartwatches: donut chart and radial bar chart are the most popular ones on squared and round-shaped screens, respectively}
        }\label{fig:my_label}
    \end{minipage}    
    \setlength{\belowcaptionskip}{-1.6cm}
\end{figure}

\begin{figure*}[h]
    \centering
    \setlength{\abovecaptionskip}{0.05cm}
    \includegraphics[width=0.95\linewidth]{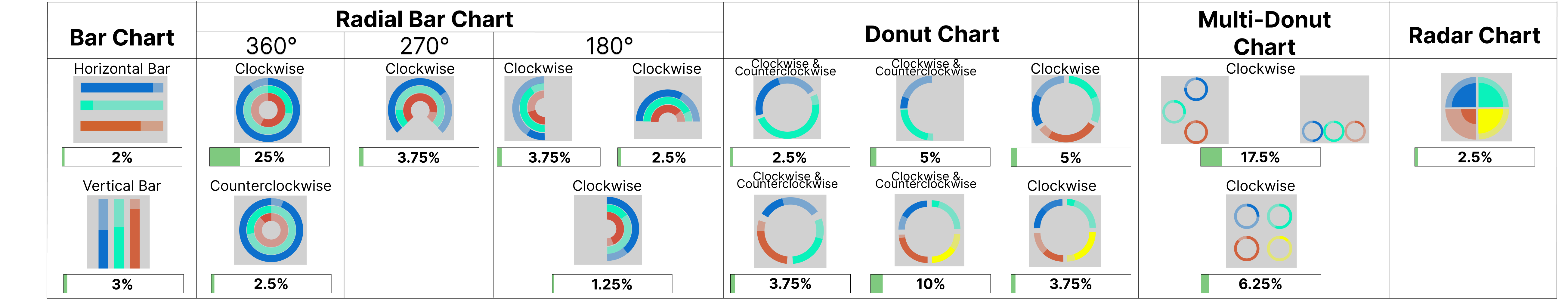}
    \caption{\emph{Activity data visualizations on smartwatches classified into five categories (bar chart, radial bar chart, donut chart, multi-donut chart and radar chart) and displayed by data volume, data arrangement and by type of data, i.e. stand, move, exercise. The progress bar under each visualization is the percentage of the share in the market.}}
    \label{fig:viztypes}
   \setlength{\belowcaptionskip}{-0.2cm}
   \vspace{-0.5cm}
\end{figure*}

The 80 smartwatch models can be equally split between round- and square-shaped display. For what concerns the visualization types, we found that overall they can be sorted into the following five categories: radial bar chart (38.75$\%$), donut chart (30$\%$), multi-donut chart (23.75$\%$), bar chart (5$\%$) and radar chart (2.5$\%$); with radial bar chart being the most popular visualization on round-screen devices, and donut chart for square-shaped devices (fig.\ref{fig:my_label}).
\vspace{-0.3cm}
\vspace{-0.3cm}
Considering the arrangement of activity data, and the scope of activity data movement (fig.\ref{fig:viztypes}), the majority of smartwatches (93.75$\%$) displays 3 types of activity data and, more in particular, the radial bar chart, with three data types, clockwise arrangement, and 360° scope (25$\%$), was found to be the most popular among round-shaped devices, followed by the multi-donut chart with 3 activities (17$\%$) among square-shaped devices.

\vspace{-1mm}

\section{QUESTIONNAIRE ON ACTIVITY DATA VISUALIZATION}
To find out more about how smartwatch wearers use activity visualizations, and to verify whether they are satisfied with the current activity visualizations, we designed and deployed an online questionnaire, which we administered using LimeSurvey.

\textbf{Questionnaire}. The questionnaire consisted in a background-questions section aimed at screening the participants' ethnography and smartwatch usage habits (e.g. brand, screen-shape, use frequency) and a section featuring the core questions about visualization preferences, which we designed by taking inspiration from the results of the exploratory study. The goal of the core section questions was to find out what users seek in activity visualizations, how they use the visualizations proposed by their activity tracking apps installed on their smartwatch and which visualizations are deemed the most effective to display activity data.

\textbf{Participants}. We recruited 74 participants through convenience sampling. 33 of them turned out not being smartwatch users, therefore we excluded them from our analysis. The remaining 41 participants (20 male, 21 female) belonged to an age range between 18 and 50 years old (AVG = 24.97, SD = 8.9) and had various occupations (25 undergraduate students, 2 postgraduate students, 2 Ph.D students, 12 working professionals).

\begin{figure}[h] \centering
\setlength{\abovecaptionskip}{-0.2cm}
\subfigure[] {
\label{fig:intentions}
\includegraphics[width=0.35\columnwidth]{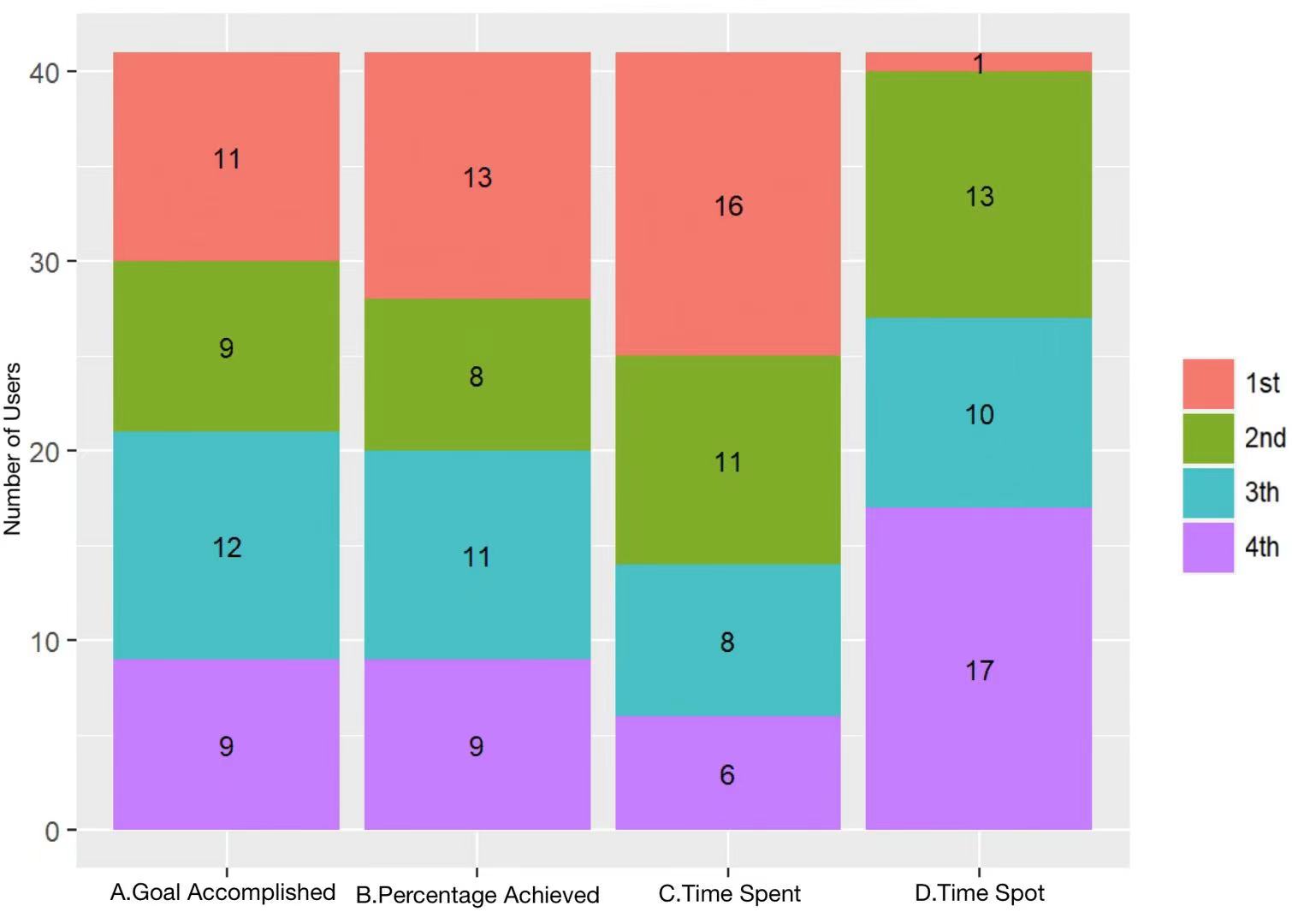}
}
\subfigure[] {
\label{fig:preferences}
\includegraphics[width=0.35\columnwidth]{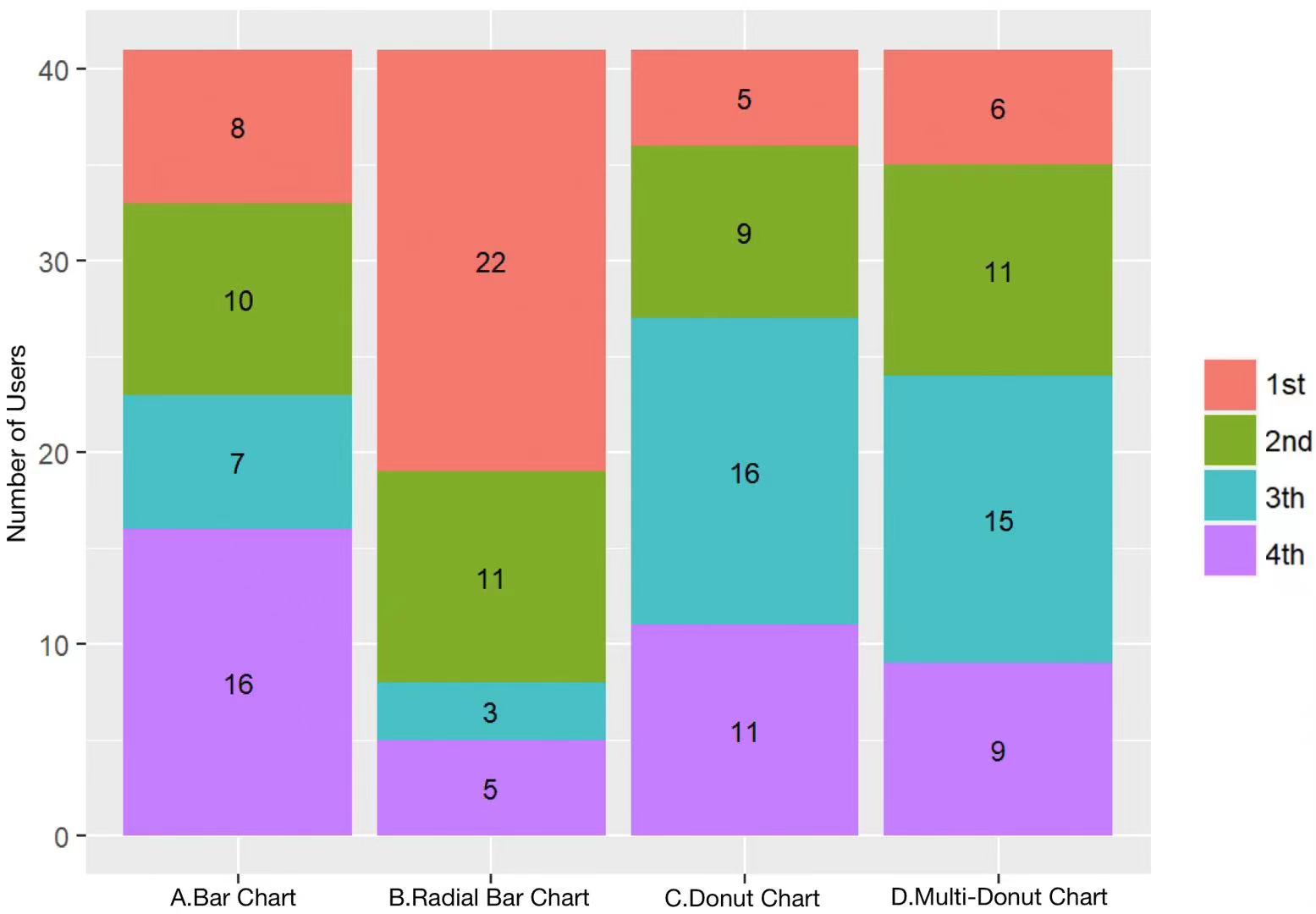}
}   
\caption{\emph{Ranking of the four intentions and four charts: \ref{fig:intentions} user interests with regards to the visualizations: time spent standing, moving, exercising (c) is the most favoured option; \ref{fig:preferences} user preferences for data visualization: radial bar chart (b) is the most preferred visualization type. 1st means the chosen choice is the most preferred. The more red the bar for an option contains, the more users consider it to be just the most important of all options. }} 
\label{fig}   
\vspace{-0.2cm}
\end{figure}

\textbf{Results and Analysis}. Starting from the background questions, we found that 20 participants reported daily smartwatch usage, 5 use their watch a few times per week, 7 a few times per month, and 9 reported a usage less than a few times per month. 
Concerning the favoured device for checking fitness data, 17 participants reported checking it on their smartwatches, 7 on their phones, and 17 on both platforms. 
The majority of participants (N=24) reported checking their fitness data after exercising, while 11 participants reported checking it when receiving a notification from their tracker, and 12 participants reported checking it during their free time during the day. 
When asked about whether they set goals for their daily activity, the majority of participants (N=23) stated that they do set goals, 15 reported not setting any goals, and 3 mentioned that their smartwatches did not support goal setting.
Concerning how the data is presented on the smartwatch, a quarter of participants (N=18) reported paying attention to both the text and visual representations when reviewing their fitness data. With regards to the visualization form for activity data offered by the participants' smartwatches, 21 chose the radial bar chart, followed by the bar chart (N=11), donut chart (N=7), and multi-donut chart (N=1).
Concerning the users' intentions about their activity (i.e. stand, move, exercise), participants were asked to rate four options according to their personal interest, i.e. ``What are you most interested in regarding your move, exercise, and stand data?''. Possible answers to the question were (A) ``Whether I accomplish the goal or not.'', (B) ``What percentage of my goals have I achieved in move, exercise, and stand?'', (C) ``How much time do I spend on move, exercise, and stand?'' and (D) ``When I move, exercise, and stand, and how long does it last?''.
The results (fig. \ref{fig:intentions}) showed that the majority of participants ($C_n$=16) ranked option C as the most important, followed by option B ($B_n$=13), option A ($A_n$=11), and option D ($D_n$=1).
Finally, the participants were asked to select the preferred representation for displaying the data about the three activities (stand, move, exercise) among four possible answers: 
We also have a question, aimed to determine the preferred representation of Move, Exercise and Stand data on the smartwatch among the following options:
(A) bar chart, (B) radial bar chart, (C) donut chart, and (D) multi-donut chart. From the results (fig. \ref{fig:preferences}) it emerged that the majority of participants ($B_n$=22) preferred radial bar charts, followed by bar charts ($A_n$=8), multi-donut charts ($D_n$=6), and donut charts ($C_n$=5). 

\setlength{\belowcaptionskip}{-0.6cm}

\vspace{-0.2cm}

\section{DISCUSSION AND CONCLUSION}
We investigated the visualizations of activity data on 80 of the best-selling smartwatches currently available on the Chinese market. We found that both square and round dials are common (40--40), but the visualizations vary across the different shapes of the dials. The most common visualization for square dials was radial bar chart; for round dials the most common was donut chart. Also the dials typically displayed 3 activities in the default state. In the state questionnaire we can see that users are interested in viewing activity data on their smartwatches. What they are mostly looking forward to knowing is the time spent on each activity followed by the percentage of goals completed. However, when investigating preferences, we found that users most preferred visualization is the radial bar chart, which is also the most common visualization normally proposed by commercially available smartwatches. This leads us to speculate that the radial bar chart is not necessarily the most effective representation for activity data, rather the high familiarity users have with this chart may have influenced their preferences to some extent.

These findings provide at least two possible directions for future research. On the one hand, it would be worth to investigate whether the perceived effectiveness of different visualizations is influenced by the shape of the smartwatch display. On the other hand, the effectiveness of the radial bar chart to display the time spent for each activity should be further investigated to find out whether it really outperforms other visualizations.

\vspace{-2mm}

\acknowledgments{
This work was supported by a XJTLU grant RDF-22-01-092.}

\vspace{-2mm}

\bibliographystyle{abbrv-doi}

\end{document}